\documentstyle[11pt]{article}
\begin{document}

\newcommand{\be}{\begin{equation}} \newcommand{\fe}{\end{equation}}
\newcommand{\eqn}{\label}\newcommand{\bel}{\begin{equation}\label}

\def\thf{\baselineskip=\normalbaselineskip\multiply\baselineskip
by 7\divide\baselineskip by 6}
\def\fff{\baselineskip=\normalbaselineskip}

\def\spose#1{\hbox to 0pt{#1\hss}}\def\lta{\mathrel{\spose{\lower 3pt\hbox
{$\mathchar"218$}}\raise 2.0pt\hbox{$\mathchar"13C$}}}  \def\gta{\mathrel
{\spose{\lower 3pt\hbox{$\mathchar"218$}}\raise 2.0pt\hbox{$\mathchar"13E$}}} 

\def\Libra{\spose {--} {\cal L}}
\def\Diam{\spose {\raise 0.3pt\hbox{+}} {\diamondsuit}  }

\font\fiverm=cmr5


\def\ov{\overline} \def\oj{\bar j} \def\hj{\hat j} \def\Ke{K}  \def\Ce{C}
\def\oT{\bar T} \def\hovT{\widehat T} \def\toT{\widetilde T}

\def\oLag{\bar L} \def\toLag{\widetilde L}

 \def\loga{\widehat{ l}\ }   \def\Lambde{\Lambda}
\def\bg{g} \def\hg{\gamma}  \def\bA{A}  \def\bF{F}  
\def\ii{i} \def\ij{j}\def\ih{h} \def\ik{k}  
\def\sigme{\sigma} \def\tor{\tau}

\def\qq{q} \def\mas{m}  \def\tmas{\widetilde m} 
\def\rast{\delta_{\!\ast}} \def\Rast{\Delta} 
\def\ortr{ r_{\!\!_\perp}\!}

\def\hovA{\widehat A}  \def\extA{\widetilde A}
\def\hovF{\widehat F}  \def\extF{\widetilde F}

\def\advr{{_{_+}}\! r} \def\retr{r}  \def\tildr{\tilde r} 
\def\Sig{{\mit\Sigma}} 

\def\ue{u} \def\ve{\tilde u} \def\acc{a} 
\def\ce{c} \def\chie{\chi} \def\phie{\varphi} \def\pe{p} \def\pee{\bar p}
\def\og{\eta} \def\ag{\perp\!} \def\nabl{\nabla\!} \def\onab{\ov\nabla\!}
\def\calK{{\cal K}} \def\tilK{\tilde{\cal K}}
\def\calE{{\cal E}} \def\psie{\psi}
\def\Ne{N} \def\pie{\pi} \def\she{ \sqrt{\kappa_{_0}}\, }

\def\Green{{\cal G}}

\title{ELECTROMAGNETIC SELF INTERACTION IN STRINGS}

\author{ {\bf Brandon Carter}
\\ D.A.R.C., (UPR 176, CNRS),
\\ Observatoire de Paris, 92 Meudon, France.}

\date{ 12 March, 1966} 

\maketitle

{\bf Abstract.} To facilitate the treatment of electromagnetic effects
in applications such as dynamically perturbed vortons, this work
employs a covariantly formulated string-source Green measure to obtain
a coherent relativistic scheme for describing the self interaction of
electromagnetic currents in string models of a very general kind, at
leading order in the relevant field gradients, using a regularised
gradient operator given by $\widehat\nabla\!_\nu=$ $\overline\nabla\!_\nu
+{1\over 2}\Ke_\nu$ where $\overline\nabla\!_\nu$ is the usual tangential 
gradient operator and $\Ke_\nu$ is the extrinsic curvature vector. 

\section{ Introduction.}

While other kinds of application can be envisaged, the main motivation for
most of the work on the development of a relativistic description of
electromagnetic self interaction in string models arises from Witten's
observation \cite{Witten85} that currents are likely to occur naturally in
many conceivable kinds of cosmic strings arising from vortex defects of the
vacuum in plausible field theories.

In the first years after Witten's epoch making observation, the importance of
electromagnetic effects was actually overestimated in comparison with that of
the purely mechanical effects of the current, which were commonly
neglected\cite{Spergel_et89}. The fact that the latter would typically be far
more important was not generally recognised until after it was shown by Davis
and Shellard that ``vortons'', meaning equilibrium states of string loops,
could easily be sustained by the purely mechanical centrifugal effect of a
circulating current\cite{DavisShellard89}, whereas the magnetostatic support
mechanism that had previously been envisaged\cite{Copeland_et88} would usually
be too weak.  This realisation lead to the development of a more realistic
kind of simplification in which electromagnetic effects (other than those
arising from a strong but purely external background field) were neglected
while attention was concentrated on the suitably non-linear treatment of the
purely mechanical effects of the current\cite{Carter89}\cite{Carter90}.

With adequate means now available for treating the primary mechanical effects
of currents in cosmic strings, it is reasonable to reconsider the secondary,
but not always negligible, effects due to electromagnetic interaction. A first
step towards the inclusion of such effects in a mechanically realistic
framework was taken by Peter\cite{Peter93}, in a study of circular vorton states.
This work confirmed the relative negligibility of the purely magnetostatic
support mechanism -- due to a spacelike current -- that had been considered
previously, but on the other hand it also showed that a potentially much more
important kind of ``spring'' effect can be produced by the electrostatic
effect of a timelike current. 

The purpose of the present work is to provide the machinery needed for
generalising Peter's work\cite{Peter93} from strictly stationary states to
dynamically perturbed configurations. It will be shown that the kind of ``core
and sheath'' model introduced in early work\cite{Copeland_et88} can be
obtained in a very different manner using the regularisation scheme developed
here, which places it on a firmer footing and shows how it can be adapted in
such a way as to be fully compatible with the more realistic mechanical models
that are now available\cite{CarterPeter95} or that may be introduced in the
future to deal for example with string phenomena in which several independent
currents may be involved\cite{Carter94}. 

The present treatment deals only with the leading order in the relevant
field gradients. This means that it does not incorporate radiation backreaction,
whose inclusion would require extending the scheme to the next differential
order, which is left for future work.

\section{ Overview of the problem.}

The primary issue dealt with here is the evaluation of the right hand side of
the two dimensional analogue of the familiar relativistic generalisation of
Newtons ``second law'' of motion for a point particle of mass $\mas$ and charge
$\qq$, which is given in terms of the unit tangent vector $\ue^\mu$ of the
worldline by
 \be\mas \acc_\mu=\qq \bF_{\mu\nu}u^\nu\, ,\hskip 1 cm
 \acc^\mu=\ue^\nu\nabl_\nu\ue^\mu \, ,  \eqn{0.0}\fe 
for an electromagnetic field $\bF_{\mu\nu}=2\nabl_{[\mu}\bA_{\nu]}$, where
the gravitational field is allowed for via the connection that specifies the
Riemannian covariant differentiation operator $\nabl$.

In the case of a string with worldsheet stress energy tensor $\oT{^{\mu\nu}}$
and electric surface current $\oj{^\nu}$, the corresponding the general
purpose ``second law'' of extrinsic motion that governs the evolution of its
worldsheet is given\cite{Carter89}\cite{Carter95} by
 \be\oT{^{\mu\nu}}\Ke_{\mu\nu}^{\,\ \ \rho}=\ag^{\!\rho\mu}
 \bF_{\mu\nu}\oj{^\nu}  \ ,  \eqn{0.1}\fe 
where  $\Ke_{\mu\nu}^{\,\ \ \rho}$ is the second fundamental tensor of the
worldsheet, and $\ag^{\!\rho}_{\,\mu}$ is the tensor of orthogonal projection .
The latter is defined in terms of the complementary tensor $\og^{\rho}_{\ 
\mu}$ of tangential projection, i.e. the first fundamental tensor, simply by
$\og^{\rho}_{\ \mu} + \ag^{\!\rho}_{\,\mu} = \bg^\rho_{\ \mu}$, 
where $\bg_{\mu\nu}$ is the background spacetime metric.  The first
fundamental tensor itself is definable as the square $\og^{\rho}_{\ \mu}=
\calE^\rho_{\ \nu}\calE^\nu_{\ \mu}$ of the antisymmetric alternating tensor
which is specifiable by the formula $\calE^{\mu\nu}=
2\ue^{[\mu}\ve^{\nu]}$ in terms of (but -- except for its orientation --
independently of the choice of) an orthonormal worldsheet tangential diad
consisting of a timelike unit vector, $\ue^\mu$ say, and the dually related
spacelike vector unit vector $\ve^\mu$, as characterised
by $\ue^\nu \ue_\nu=-1$, $\ve^\nu\ve_\nu=1$, $\ue^\nu\ve_\nu=0$. The second
fundamental tensor is defined in terms of the first one by
 \be\Ke_{\mu\nu}^{\,\ \ \rho}=\og^\sigma_{\ \nu}\onab_\mu\og^\rho_{\ \sigma}
 \ , \hskip 1 cm \onab_\mu=\og^\nu_{\ \mu}\nabl_\nu\ .\eqn{0.2}\fe

As the natural generalisation to two dimensions of the familiar ``second
law'' (\ref{0.0}) the extrinsic dynamical equation (\ref{0.1}) should in
principle be applicable -- in the thin limit when higher order
longitudinal derivatives are negligible -- for any kind of string
model, including even the relatively complicated kind that would be
needed to describe a warm, dissipatively conducting terrestrial power
cable.  Like its 1-dimensional analogue (\ref{0.0}), the 2-dimensional
equation of motion (\ref{0.1}) is straightforwardly applicable so long
as the field $\bF_{\mu\nu}$ can be considered to be of purely external
origin. However, again like (\ref{0.0}), the equation (\ref{0.1}) ceases
to be so obviously meaningful when one notices that although the locally
source free external contribution, $\extF_{\mu\nu}$ say, will be well behaved,
the  self induced contribution $\hovF_{\mu\nu}$ to the total $\bF_{\mu\nu}=
\hovF_{\mu\nu}+\extF_{\mu\nu}$ will be divergent in the zero thickness limit.

For a realistic treatment one must recognise that, in a high resolution
description, the underlying physical system (which for a cosmic string will be
some kind of  vortex defect of the vacuum) will in fact have a finite
effective thickness. This will provide a natural ``ultraviolet'' cut off
radius, $\rast$ say, for a scheme whereby the regularised value of any field
is covariantly specified as the average of its values at a distance $\rast$
in directions orthogonal to the world sheet. 

In the point particle case the regularised field tensor $\hovF_{\mu\nu}$ will
have the form
 \be \hovF_{\mu\nu}={\qq\over\rast} \acc_{[\mu}\ue_{\nu]}\ ,\eqn{0.4}\fe
in which -- as throughout the present discussion --
we are neglecting the higher derivative contributions that would be
needed to take account of radiation backreaction. If so desired,
the self interaction force due to (\ref{0.4}) in (\ref{0.0})  can be
allowed for by transferring the relevant term from the right to the
left and then absorbing it into the first term by a mass
renormalisation of the usual kind $\mas \mapsto \tmas$ where
 \be\tmas=\mas + {\qq^2\over 2 \rast} \, .\eqn{0.5}\fe

Our purpose here is to carry out the analogous regularisation procedure -- and
thus to provide the option of a corresponding renormalisation if so desired --
for the two dimensional case of a string. In this case the leading
``ultraviolet'' divergence is not linear but only logarithmic, so that as well
as the radius $\rast$ characterising the effective radius of the string (or to
be more specific, that of its current distribution) one also needs a
macroscopic ``infrared'' cut off length, $\Rast$ say. It fortunately turns out
that no anomally arises in the leading order ``ultraviolet'' divergence
considered here, despite the fact that there is not any way of actually
performing the ``infrared'' cut off in a strictly covariant manner. 
(It will not be so easy to avoid trouble with Lorentz invariance at the higher
order that would be needed for treating radiation backreaction.)

\section{Overview of the solution.}

The elegantly covariant result one obtains at leading order is as follows.  
To start with -- as was remarked at the outset by Witten\cite{Witten85}, and
will be made obvious below -- the leading contribution to the regularised self
field $\hovA_\mu$  on the string world sheet will be expressible (using an
appropriate gauge) in terms of of the surface current $\oj{^\mu}$ there by a
simple proportionality relation of the form
 \be \hovA_\mu =\loga\oj_\mu\eqn{0.6}\fe
in which $\loga$ is a dimensionless constant of the familiar form 
 \be \loga=2\ln\big\{ {\Rast/\rast} \big\}\, .\eqn{0.7}\fe
    
The mathematically non-trivial part of the problem is the derivation of
the corresponding regularised field $\hovF_{\mu\nu}$.  The worldsheet
tangential part can be obtained in the usual way by derivation of $\hovA_\mu$,
but since the regularised value is defined only on the worldsheet itself, it
cannot be directly differentiated in non tangential directions:
$\onab_\nu\hovA_\mu$ is directly meaningful, but $\nabl_\nu\hovA_\mu$ is not.
One can however obtain the required result by going through the regularisation
procedure again at first differentiated order. 

The final result is most conveniently expressible in terms of the self-dual
part
 \be \Ce_{\mu\nu}^{\,\ \ \rho}={_1\over^2}\big(
 \Ke_{\mu\nu}^{\,\ \ \rho}+\tilK_{\mu\nu}^{\,\ \ \rho}\big) \fe
of the second fundamental tensor, where the dual is defined by
 \be \tilK_{\mu\nu}{^\rho}=
 \calE_{\mu\sigma}\calE_{\nu\tau}\Ke^{\sigma\tau\rho}=\Ke_{\mu\nu}{^\rho}
 -\og_{\mu\nu}\Ke^\rho   \, ,\fe
 in which $\Ke^\rho$ is the trace vector,
 \be\Ke^\rho=\Ke_\mu^{\ \mu\rho}=\onab_\mu\og^{\mu\rho}  \eqn{0.9} \fe
(whose vanishing, $K^\mu=0$, would express the dynamical equations of motion
for a string model of the simplest Goto-Nambu type in the absence of any
external force or self interaction). The tensor $\Ce_{\mu\nu}^{\,\
\ \rho}$ is identifiable as the conformally invariant {\it conformation
tensor} whose original definition\cite{Carter92}, 
 $ \Ke_{\mu\nu}^{\,\ \ \rho}=\Ce_{\mu\nu}^{\,\ \ \rho}+{_1\over ^2}
 \og_{\mu\nu}K^\rho$,
shows that it is trace free, $\Ce_{\mu}^{\,\ \mu\rho}=0$.

Although $\Ke^\rho$ will be usually be non zero in electromagnetically
interacting string models, its contribution cancels out in the final formula
for the self field, which reduces, as shown below, simply to
 \be \hovF_{\mu\nu}=2\og_{\,[\mu}{^\rho}\onab_{\nu]}\hovA_\rho
 +2\hovA^\rho \Ce_{\rho[\mu\nu]} \, .\eqn{0.10}\fe

\section{Retarded Green measure for a string.}

To carry out the first differential order treatment given here (not to mention
the second differential order treatment that will need to be carried out in
future work to allow for radiation backreaction) the essential tool is the
appropriate Green formula for the relevant retarded solution. Since we are
concerned here only with the short range ``ultraviolet'' divergence
contribution it will suffice to use the flat space Green formula.
(For our present purpose it would make no difference if we used
the advanced Green formula: it is only at the next higher differential order
that the distinction becomes important.)  

For an an isolated particle with charge $\qq$ the field will be given 
simply by
\be \bA^\mu={\qq\ue^\mu\over\retr^\nu \ue_\nu}\, , \eqn{1.4}\fe
where $\ue^\mu$ is the future directed unit tangent vector to the trajectory
at the (unique) point where it is intersected by the past null cone from the
point of interest, and $\retr^\mu$ is the past directed null vector
-- as characterised by $\retr^\nu\retr_\nu=0$, $\retr^\nu \ue_\nu>0$ --
from this ``observation position'' to the intersection point on the worldline.
What we need is the analogous formula for a string, as provided in the
appropriate limit by the well known Li\'enard-Wiechert formula which solves
the source equation in the hyperbolic version 
 \be\nabl_\nu\nabl^{\,\nu} A^\mu=-4\pi\hj^\mu\, ,\eqn{1.5}\fe 
(obtained from the gauge invariant version $\nabl_\nu
F^{\mu\nu}=4\pi\hj^\mu$ by imposition of the Lorentz condition $\nabl_\nu
A^\nu=0$) for a smooth current distribution $\hj{^\mu}$, in a background with
Minkowski metric $ ds^2=$ $(dx^{_1})^2+(dx^{_2})^2+(dx^{_3})^2-(dx^{_0})^2$,
in the form
 \be\bA^\mu=\int\int\int  \hj{^\mu}
 \, { dx^{_1}\, dx^{_2}\, dx^{_3}\over \retr_{_0} }\, . \eqn{1.6}\fe
(Though not manifestly covariant, this formula does in fact give a well
defined Lorentz covariant result for a suitably bounded source 
distribution.)

For the limiting case of a distribution confined to a string, with surface
current density $\oj{^\nu}$ that is tangential to the worldsheet, the result
will evidently be obtainable as an integral along the one dimensional curve
where the relevant (past or future) null cone from the point under
consideration intersects the worldsheet.  With respect to a suitably
normalised parameter $\Green$ this curve will have a tangent vector given by
 \be d\bar x^\mu=\tildr^\mu\, d\Green\, , \hskip 1 cm
 \tildr^\mu=\calE^{\mu\nu}r_\nu\, ,\eqn{1.10}\fe 
where, as before, $\retr^\mu=\bar x^\mu-x^\mu$  is the {\it retarded
null vector} specifying the ray that goes from the external observation
position $x^\mu$ to the relevant point $\bar x^\mu$ on the worldsheet.
It turns out that the dimensionless parameter $\Green$ introduced in this way
directly specifies the relevant {\it Green measure}, in terms of which 
the required analogue of (\ref{1.4}) giving the contribution
from a finite segment with beginning and end labeled by $-$ and $+$
will be given simply by 
 \be\bA^\mu=\int_-^+  \oj{^\mu}\, {d\Green}\, .\eqn{1.7}\fe
The Green measure given by (\ref{1.10}) can be seen to be related to the
ordinary  positive indefinite {\it proper} length measure $ds$ along the
curve by
\be d\Green={ds\over\ortr} \, ,\hskip 1 cm \ortr
 =\sqrt{\tildr^\mu\tildr_\mu} =\sqrt{\ag^{\!\mu\nu}\retr_\mu\retr_\nu}
\, .\eqn{1.8}\fe
In terms of worldsheet coordinates $\sigme^\ii$ and of the corresponding
2-dimensional intrinsic metric given by $\hg_{\ii\ij}= g_{\mu\nu}\bar
x^\mu{_{,\ii}} \bar x^\nu{_{,\ij}}$ (using a comma to denote partial
differentiation) the fundamental tensor will be given simply by $\og^{\mu\nu}
=\hg^{\ii\ij}\bar x^\mu{_{,\ii}} \bar x^\nu{_{,\ij}}$, so 
this works out as $d\Green=$ $\vert\hg^{\ii\ij}\bar
x^\nu{\!_{,\ii}} \bar x^\rho{\!_{,\ij}}\retr_\nu \retr_\rho\vert^{-1/2} 
\sqrt{\hg_{\ih\ik}\, d\sigme^\ih\, d\sigme^\ik }$.

Unlike (\ref{1.10}), the explicit formula (\ref{1.8}) becomes
indeterminate wherever the tangent direction $\tildr^\mu$ is null so
that the orthogonally projected distance $\ortr$ vanishes. This can be
overcome at the expense of manifest covariance by referring to a
worldsheet frame consisting of any future directed timelike unit
tangent vector $\ue^\mu$, and the dually related spacelike unit vector
$\ve^\mu= \calE^{\mu\nu}\ue_\nu$, which gives the everywhere well
behaved expression $d\Green= (u_\nu r^\nu)^{-1}\ve_\mu d\bar x^\mu$.

What is commonly quoted in the literature is a more specialised
version\cite{Copeland_et90} that is given by $d\Green= (\dot x^\nu r_\nu)^{-1}
\, d\sigme$, but that is valid only for coordinates $\sigme$, $\tau$ of {\it
conformal} type (meaning that $\dot x^\nu\dot x_{\,\nu}
+x^{\prime\nu}x^\prime_{\,\nu}=0$, $\dot x^\nu x^\prime_{\,\nu}=0$, where
$\dot x^\nu=\partial \bar x^\nu/\partial\tor$, $x^{\prime\nu}=\partial \bar
x^\nu/\partial\sigme$).

\section{ Regularised gradient operator for self field on worldsheet.}

It is almost immediately evident that the leading order contribution 
to the regularised effective gauge potential on the string will be given 
by an expression of the form (\ref{0.6}). The non-trivial part of the
calculation is the evaluation the corresponding effective gradient components
as analogously regularised by averaging over an infinitesimal cirle on which
$\ortr=\rast$. The terms that remain at leading order give a result of the
form
 \be\widehat{\nabl_\nu A}{^\mu}= \loga\Big(  \onab_\nu\oj{^\mu}
 +{_1\over ^2}\Ke_\nu \oj{^\mu}\Big)\, ,\eqn{2.6}\fe using the same
regularisation coefficient $\loga$ as was introduced in (\ref{0.6}).
For the tangentially differentiated components in the first term, the
postulate that this coefficient should be constant ensures that the
result of the regularisation procedure will be consistent with what one
would get by direct tangential differentiation of the regularised
components in the world sheet. This means  that (\ref{2.6}) will be
expressible by $\widehat{\nabl_\nu A{^\mu}}= \widehat\nabl_\nu
\hovA{^\mu}$ in terms of the {\it regularised gradient operator} given
by \be\widehat\nabl_\nu= \onab_\nu+ {_1\over ^2}\Ke_\nu \, .\eqn{2.14}\fe
With this notation the required self interaction field will be
given simply by
 \be\hovF_{\mu\nu}=2\widehat\nabl_{[\mu}\hovA_{\nu]} \ .\eqn{2.15}\fe
Reorganising this  regularised effective field in terms of its distinct purely
tangential and mixed components (it has no purely orthogonal one) using
the tangentiality property $\ag^{\!\mu}_{\, \nu}\hovA^\nu=0$ implied by
(\ref{0.6}), which entails the identity
$\ag^{\!\rho}_{\,\nu}\onab_\mu\hovA{^\nu}=$ $\hovA^\nu \Ke_{\mu\nu}^{\, \ \
\rho}$, one finally obtains the quoted formula (\ref{0.10}).

\section{ Core and sheath representation.}

Having obtained the main result (\ref{2.15}), or equivalently (\ref{0.10}),
that was required, one naturally hopes to find that, as in the point particle
case, the effect of the ensuing self force contribution will be absorbable
into the left hand side of the ``second law'' equation (\ref{0.1}) by the use
of an appropriate two dimensional analogue of the renormalisation (\ref{0.5}).
The concept of such an adjustment was already introduced in the early work of
Copeland, Haws, Hindmarsh, and Turok\cite{Copeland_et88} who developped a
treatment whereby the current carrying cosmic string of strictly confined
``local type'' was considered as forming the core of a composite string with
an outer sheath of the extended ``global type'' constituted by its own
electromagnetic field. In a compound  ``core and sheath'' string model of this
kind, the outer electromagnetic sheath provides an extra contribution to the
effective tension and energy per unit length that will be representable by a
worldsheet stress energy tensor of the form
 \be\hovT{^{\mu\nu}}=\loga\big(\oj{^\mu} \oj{^\nu}- {_1\over ^2}\oj{^\rho}
 \oj_\rho\og^{\mu\nu} \big) \ ,\eqn{1.2}\fe 
where $\loga$ is a dimensionless regularisation coefficient of the same form
(\ref{0.7}) as the one introduced above. Despite their mathematical similarity
at the end, it is to be noticed that the physical origin of the coefficient in
(\ref{0.6}) was very different from that of the one in (\ref{1.2}), which came
from an integral over a section through the field {\it outside} the current
distribution, whereas the $\loga$ in (\ref{0.6}) comes from an integral with
support confined to the current carrying core. 

In view of this distinction, it is remarkable that the two approaches turn out
in the end to agree perfectly. The form of the self field (\ref{0.10}) is
easily be seen to be such that the corresponding self force contribution in
(\ref{0.1})  can indeed be taken over to the left hand side and absorbed into
the first term by an adjustment of the form $\oT{^{\mu\nu}}
\mapsto\toT{^{\mu\nu}}$, in which the required two dimensional analogue of the
mass renormalisation (\ref{0.5}) turns out to have precisely the form that
describes the total in the compound core and sheath model, namely 
  \be\toT{^{\mu\nu}}=\oT{^{\mu\nu}}+\hovT{^{\mu\nu}} \, ,\eqn{1.3}\fe 
where the constant coefficient $\loga$ in (\ref{1.2}) is to be identified
precisely with the constant coefficient $\loga$ in (\ref{0.6}).

This demonstration that it agrees with the result of the more refined analysis
presented here puts the compound ``core and sheath'' type of model on a firmer
footing than hitherto. The agreement is not limited to the prediction of the
extrinsic equation of motion discussed above, but still holds just as well for
the internal dynamics of the two kinds of model. To see this, let us consider
the dynamical evolution of the  loop in an electromagnetic background that may
include a locally source free external field contribution $\extF_{\mu\nu}=$
$\nabl_{[\mu}\extA_{\nu]}$ as well as the self field contribution
(\ref{0.10}), so that the relevant effective total to be substituted in
(\ref{0.1}) is 
\be \bF_{\mu\nu}=\hovF_{\mu\nu}+ \extF_{\mu\nu}\, .\eqn{3.1}\fe
The extrinsic dynamical equation (\ref{0.1}) is not sufficient
by itself to determine the evolution of the string configuration:
it is also necessary to know the equations of motion of the surface current
$\oj{^\mu}$ and any other independent internal fields on which the worldsheet
stress energy tensor $\oT{^{\mu\nu}}$ may depend. Whatever the
nature of the internal fields, the internal momentum-energy
transport equation
 \be \og^\rho_{\ \mu}\nabl_\nu\oT{^{\mu\nu}}=\og^{\rho\mu}\bF_{\mu\nu}
 \oj{^\nu}  \eqn{3.2}\fe
must be satisfied.
Writing the ``sheath'' contribution (\ref{1.2})  in the form
 \be\hovT{^\nu}_\mu=\oj{^\nu}\hovA_\mu- {_1\over ^2}
 \oj{^\rho}\hovA_\rho \og^{\nu}_{\ \mu} \, ,\eqn{3.3}\fe 
one sees that, subject to the surface current conservation condition
 \be \onab_\nu \oj{^\nu}=0 \, , \eqn{3.4}\fe
the self force contribution on the right in (\ref{3.2}) can also be taken over
to the left and absorbed into the first term by the same adjustment
$\oT{^{\mu\nu}} \mapsto \toT{^{\mu\nu}}$ as before, where $\toT{^{\mu\nu}}$
is the total for the compound core and sheath model as given by (\ref{1.3}).
After performing such a transfer both for the extrinsic dynamical
equation (\ref{0.1}) and the extrinsic dynamical equation (\ref{3.2})
the results can be combined in a total force law of the form
 \be \onab_\nu\toT{^\nu}_{\!\mu}= \extF_{\mu\nu}\oj{^\nu}\, ,\eqn{3.5}\fe
in which only the locally source free purely external field contribution
$\extF_{\mu\nu}$ is involved on the right hand side.

\section{ Action formulation for simply conducting case.}

What has done so far is valid even for very general string models involving
multiple conductivity and finite resistivity as in terrestrial power cables.
Let us now restrict attention to the simple strictly conservative kind of
model that is appropriate for a Witten type superconducting string, in which
the only independent dynamical variable can be taken to be a particle current
vector $\ce^\mu$ say, in terms of which the electromagnetic current will be
specifiable by a proportionality relation of the form 
\be \oj{^\mu}=\qq\,\ce^\mu\, , \eqn{2.1} \fe
where $\qq$ is the relevant mean charge per particle. Such a 
model will be specified by a {\it master function}, $\Lambde$ say, with a 
a -- typically  non-linear\cite{CarterPeter95} -- dependence on the single scalar variable
$\chie=\ce^\nu\ce_\nu$, so that its variation determines a corresponding
momentum vector given by $\delta\Lambde=\pee_\nu\delta\ce^\nu$  with
$\pee_\mu=\calK \ce_\mu$ where $\calK=2 d\Lambde/d\chie$. 
When self interaction is neglible, the mechanics will be governed just by a
Lagrangian density scalar of the simple form
 \be \oLag =\Lambde + \qq\,\ce^\nu\bA_\nu \, ,\eqn{3.6}\fe
whose variation determines a gauge dependent total momentum covector:
$\delta\oLag=\pie_\nu\delta \ce^\nu$ where 
 \be \pie_\nu=\pee_\nu+\qq\bA_\nu \, .\eqn{3.7}\fe 
In the application of the variation principle to the corresponding action
$\int \oLag\sqrt{\Vert\hg\Vert} \, d\sigme\, d\tor$, the current is not 
an entirely free variable, but must be constrained so as to be conserved,
e.g. by taking it to be dual to the gradient of a freely variable stream function $\psie$
say, i.e. $\ce^\mu=\calE^{\mu\nu}\onab_\nu\psie$. For such a
model, the worldsheet stress energy tensor reqired for application of the
extrinsic equation of motion (\ref{0.1}) takes the form
 \be \oT{^\mu}_{\nu}= \ce^\mu\pee_\nu+(\Lambde-\ce^\rho\pee_\rho)\og^\mu_{\ 
 \nu}\, . \eqn{3.8}\fe

It is easy to see that appart from the kinematic particle conservation
law $\onab_\nu\ce^\nu=0$, whose consequence (\ref{3.4}) ensures that the loop 
will be characterised by a conserved total charge,
$ Q=\oint\calE_{\mu\nu}\,\oj{^\nu}\, dx^\mu$,
the  remaining internal dynamical equations for the foregoing model will 
consist just of the surface integrability condition
 \be \og_{\,[\mu}{^\rho} \onab_{\nu]} \pie_\rho=0 \, , \eqn{3.9}\fe 
which ensures that the the tangentially projected part of the momentum
covector (\ref{3.7}) is proportional to the surface gradient of a
scalar, i.e. $ \og^{\mu\nu}\pie_\nu=\she\onab^{\,\mu}\phie$, 
where $\she$ is a fixed proportionality factor. It follows that the string
loop will be characterised by a conserved circuit integral given by
$ 2\pi\Ne=\oint d\phie$. The inclusion of the proportionality constant 
$\she$ allows the scalar $\phie$ to be adjusted\cite{CarterPeter95} in 
order to agree with the phase of a boson condensate in an underlying 
microscopic model, so that it will be periodic with period $2\pi$, in 
which case $\Ne$ will be quantised as an integral winding number.

It can be seen that when the self interaction is taken into account using the
regularisation scheme given above, the foregoing system of equations will
still apply provided the field $\bA_\mu$ in the specification (\ref{3.7}) of
the total momentum per particle is interpreted in accordance with (\ref{3.1})
as the combination $\bA_\mu=\hovA_\mu+\extA_\mu$, in which $\extA_\mu$ is the
well behaved source free external contribution and $\hovA_\mu$ is the
regularised self field as given by (\ref{0.6}).

The same equations of motion can be obtained by a purely variational approach
(something that would not be possible at the next differential order when
radiation backreaction will be involved) within the framework of the ``core
and sheath''  model, whose dynamics will be expressible in variational form
using an appropriately modified Lagrangian given by $\toLag= \widetilde\Lambde
+\qq\ce{^\nu}\extA_\nu$ in which only the external field contribution
$\extA_\mu$ is involved, the self interaction contribution having been
absorbed into the uncoupled term by an additive renormalisation $\Lambde
\mapsto \widetilde{\Lambde}$ with 
 \be\widetilde{\Lambde}=\Lambde+\widehat{\Lambde}\, , \hskip 1 cm
 \widehat{\Lambde}={_1\over ^2}\,\loga\qq^2\ce^\nu\ce_\nu\, .\eqn{3.14}\fe 
The ``sheath'' contribution, $\widehat{\Lambde}=$ ${1\over2}\,\loga\qq^2
\chie$, is interpretable as the effective action density of the self generated
electromagnetic field, as evaluated\cite{Copeland_et88} by integration across
an external section. The effective momentum per particle will undergo a
corresponding adjustment  $\pee_\mu\mapsto \widetilde\pe_\mu$ with
$\widetilde\pe_\mu= \pee_\mu+\widehat\pe_\mu$ where the ``sheath''
contribution is $\widehat\pe_\mu=\loga\qq^2\ce_\mu$. The corresponding gauge
dependent total, $\widetilde\pie_\mu=$ $\widetilde\pe_\mu+\qq\extA_\mu$, is
the same as was given by (\ref{3.7}), i.e. one obtains
$\widetilde\pie_\mu=\pie_\mu$, so no adjustment is needed for the phase scalar
$\phie$ or its conserved winding number $\Ne$.

I wish to thank P. Peter, M. Sakellariadou, and A. Gangui for illuminating
discussions.

\end{document}